\documentclass{PoS}

\title{Searches for supersymmetry in final states with leptons or photons and missing energy}

\ShortTitle{Search for SUSY in leptonic final states}

\author{\speaker{Sanjay Padhi}\thanks{On behalf of the CMS Collaboration}\\
        Department of Physics, University of California, San Diego\\
        E-mail: \email{Sanjay.Padhi@cern.ch}}

\abstract{We present the results of searches for Supersymmetry in various topologies that lead to final states with jets and missing transverse momentum together with one or more isolated leptons, one or two photons or a photon and a lepton. The searches are performed using data collected by the CMS experiment at the LHC in pp-collisions at a center-of-mass energy of 7 TeV. Various data-driven techniques used to measure the Standard Model backgrounds are discussed. The results are interpreted in the CMSSM framework.}

\FullConference{The 2011 Europhysics Conference on High Energy Physics-HEP 2011,\\
		July 21-27, 2011\\
		Grenoble, Rhodes Alpes France}

\begin{document}
\section{Introduction}
In this article results of searches for Supersymmetry (SUSY) using $\approx$ 1 fb$^{-1}$ of integrated luminosity 
collected by the CMS experiment~\cite{bib:cms} at the LHC in pp-collisions at a center-of-mass energy of 7 TeV are presented. CMS 
has conducted several inclusive searches, categorized by the number of leptons in the
final state in association with hadronic jets to be sensitive to the strong production and missing transverse momentum (MET). 
The discovery potential of the new physics entirely depends on the modeling of the Standard Model (SM) backgrounds. 
Several data-driven methods are developed for constraining and measuring these backgrounds directly from the data. 
Searches involving photons can be found elsewhere~\cite{bib:public}.
\section{Inclusive searches with data-driven background predictions}
Three search results involving dileptons in association with jets and MET are presented along with their respective background determination methods. The first search~\cite{bib:os} 
uses opposite sign dileptons (OS) with veto on the $Z$ boson. The second search~\cite{bib:osz1, bib:osz2} covers physics beyond the SM in final states with a Z boson. 
Finally, same-sign isolated dilepton (SS) modes~\cite{bib:ss} are explored, that are rare in the SM, but occur naturally in several SUSY signatures. 
Jets and MET are reconstructed using the Particle Flow technique. The anti-k$_T$ clustering algorithm with $\Delta$R = 0.5 has been used for jet clustering.

The SM processes that can produce OS dilepton ($ee, \mu\mu, e\mu$) signatures are predominantly $t\bar{t}, W/Z$+jets, dibosons, single top and 
QCD multijet events. Isolated electrons and muons are required to have $p_T > 10$~GeV, with the leading lepton with $p_T > 20$~GeV. Events with OS same 
flavor ($e^+e^-, \mu^+\mu^-$) with invariant mass between 76 GeV/$c^2$ and 106 GeV/$c^2$ or below 12 GeV/$c^2$ are removed, in order to suppress Drell-Yan (DY), 
as well as low mass dilepton resonances. After the pre-selection requirement on the scalar sum of the transverse energies of the jets ($H_T$) $> 100$~GeV and 
MET > 50 GeV, the dominant remaining background is from $t\bar{t}$ (81\%). Two signal regions: high MET (MET > 275 GeV, $H_T > 300$~GeV) and 
high $H_T$ (MET > 200 GeV, $H_T > 600$~GeV) are defined. The study uses three independent methods to estimate the background from data in the signal 
regions once the methods are validated using the control samples. The first method exploits the nearly uncorrelated variables $H_T$ and $y \equiv$ 
MET/$\sqrt{H_T}$ referred as ABCD' technique. The differential distributions $f(y)$ and $g(H_T)$ are extracted from the data using control regions 
in the plane of $y$ vs. $H_T$. The number of events in the signal region is predicted using double differential 
$\frac{\partial^2 N}{\partial y \partial H_T} = f(y) g(H_T)$. The bin contents of $f(y)$ and $g(H_T)$ are smeared according to their Poisson uncertainties
to estimate statistical errors. The correction factors include uncertainties due to possible bias from the small correlation between 
the variables. The background estimations in the aforesaid signal regions are given in Table~\ref{tab:os1} and are found to be consistent with 
the MC expectations as well as with the observed events in data. 
\begin{table}[hbt]
\begin{center}
\caption{\label{tab:os1} 
Summary of the observed and predicted yields in the two signal regions. The uncertainty in the MC prediction is statistical only. The non-SM yield UL is a 
hybrid frequentist-bayesian CLs 95\% confidence level upper limit based on the error-weighted average of the two data-driven predictions.}
{\footnotesize
\begin{tabular}{l|c|c}
\hline
                                       &     high MET signal region             &  high $H_T$ signal region              \\ 
\hline
observed yield                         &                          8             &                        4              \\
\hline
MC prediction                          &              7.3 $\pm$ 2.2             &            7.1 $\pm$ 2.2              \\
ABCD' prediction                       &   4.0 $\pm$ 1.0 (stat) $\pm$ 0.8 (sys)  & 4.5 $\pm$ 1.6 (stat) $\pm$ 0.9 (sys) \\
$p_T(ll)$ prediction                   &   14.3 $\pm$ 6.3 (stat) $\pm$ 5.3 (sys)  & 10.1 $\pm$ 4.2 (stat) $\pm$ 3.5 (sys) \\
non-SM yield UL                        &                 10                      &               5.3                     \\
\hline
\end{tabular}}
\end{center}
\end{table}
The second background estimation method ($p_T(ll)$) uses the fact that up to polarization effects, leptons and neutrinos from $W$ decays have
the same momentum distribution on average. Here the observed $p_T(ll)$ distribution 
is used to model the $p_T(\nu\nu)$ distribution, which is 
identified as MET. Correction factors on the $p_T(ll)$ distribution are applied to account for the MET > 50 GeV pre-selection requirement and $W$ polarization effects 
in the $t\bar{t}$ system, including the detector uncertainties. 
The predictions are consistent with the observation shown in Table~\ref{tab:os1}.
The third background estimation method is based on quantifying the excess of same flavor (SF) vs. opposite flavor (OF) dileptons pairs using 
$\Delta = R_{\mu e} N(ee) + \frac{1}{R_{\mu e}} N(\mu\mu) - N(e\mu)$. The muon to electron efficiency ratio $R_{\mu e} = 1.12 \pm 0.05$ is measured
using the $Z$ control sample. After applying the $Z$ mass veto for all lepton flavors and the OF subtraction, $\Delta = 3.6 \pm 2.9 \pm 0.4 (-0.9 \pm 1.8 \pm 1.1) $ 
for the high MET (high $H_T$) region, this is found to be consistent with no deviation from the SM. 
In addition, the search for the kinematic edge in order to be sensitive to decays such as $\tilde{\chi^{0}_{2}} \rightarrow l \tilde{l} \rightarrow l^+ l^- \tilde{\chi^{0}_{1}}$ 
in the dilepton mass distribution for same-flavor events are studied. The result of this search is consistent with background only hypothesis. 
\begin{table}[hbt]
\begin{center}
\caption{\label{tab:osz} 
Summary of the observed, predicted and MC yields for $JZB$ and MET template methods. The upper limits (UL) on signal events at 95\% C.L. are estimated for 
$JZB$ and  MET template studies using 0.19 fb$^{-1}$ and 0.98 fb$^{-1}$ of integrated luminosity, respectively.}
{\footnotesize
\begin{tabular}{l|c|c|c|c}
\hline
                                       &  Observed yield   &  Background prediction      & UL &   MC expectation      \\ 
\hline
$JZB > 50$~GeV                         &      20    &  24 $\pm$ 6 (stat) $\pm$ 1.4(peak)$^{+1.2}_{-2.4}$ (sys) & 11.6 & 16.0 $\pm$ 1.2 (MC stat) \\
$JZB > 100$~GeV                        &       6    &  8 $\pm$ 4 (stat) $\pm$ 0.1(peak)$^{+0.4}_{-0.8}$ (sys)  & 6.6 & 3.6 $\pm$ 0.4  (MC stat)    \\
\hline
MET > 30 GeV (Template)                &      2287  &  2306 $\pm$ 29.7 (stat) $\pm$ 309.9 (sys) & 498  & -         \\
MET > 60 GeV (Template)                &      206   &  213 $\pm$ 6.4 (stat) $\pm$ 16.5 (sys)   & 37 & -         \\
MET > 100 GeV (Template)               &      57    &  55.7 $\pm$ 3.0 (stat) $\pm$ 4.6 (sys)   & 20 & -         \\
MET > 200 GeV (Template)               &      4     &  3.3 $\pm$  0.7 (stat) $\pm$ 0.3 (sys)   & 5.9 & -         \\
\hline
\end{tabular} }
\end{center}
\end{table}

Searches for new physics involving $Z$ bosons with MET signatures such as: $\tilde{\chi^{0}_{2}} \rightarrow Z \tilde{\chi^{0}_{1}}$ are performed~\cite{bib:osz1, bib:osz2}. 
Two complementary
data-driven methods $JZB$~\cite{bib:osz1} and MET template~\cite{bib:osz2} methods are used to quantify the possible enhancement of events above the SM backgrounds (predominantly
composed of DY with instrumental MET and $t\bar{t}$ productions). The $t\bar{t}$ contributions are estimated using the OF subtraction method. 
The $JZB$ variable is defined as the difference between the $p_T$ of the vectorial sum of all 
jets and the $p_T$ of the $Z$ boson candidate. Events with pairs of OSSF leptons with $p_T > 20$~GeV and invariant mass within 20 GeV of the $Z$ 
boson mass are selected. At least three jets with $p_T > 30$~GeV and $|\eta| < 3.0$ are required. The backgrounds arising from $Z$ associated with instrumental MET 
are determined using $JZB < 0$ control region. Table~\ref{tab:osz}
shows the number of observed events in two signal regions with $JZB > 50, 100$~ GeV along with the estimated backgrounds from MC, as well the combination
of above mentioned methods. 
The MET template method, uses a similar event selection as the $JZB$ study, but with at least two jets and with dilepton invariant mass within 10 GeV of the $Z$ boson.
The instrumental MET backgrounds from the $Z$ production are estimated using templates of MET derived as a function of jet multiplicities and $H_T$ in 
QCD and photon plus jets samples. The effect of the differences between the distributions of hadronic recoil $p_T$ in the control vs. signal
samples is estimated by reweighting the photon plus jets events such that the hadronic recoil $p_T$ distribution matches that in the $Z$ plus jets events.
The total uncertainty of $\approx 15$\% is assigned, which includes variation in photon selection, as well as the kinematic differences in the prediction 
between the two bosons. Table~\ref{tab:osz} summarizes the comparison between observed and estimated backgrounds.
In absence of any enhancement above the SM background model independent upper limits at 95\% C.L. on the non-SM contribution to the yields are also provided in the table.
\begin{figure}[htb]
\begin{center}
\includegraphics[width=0.48\linewidth, height=0.33\linewidth]{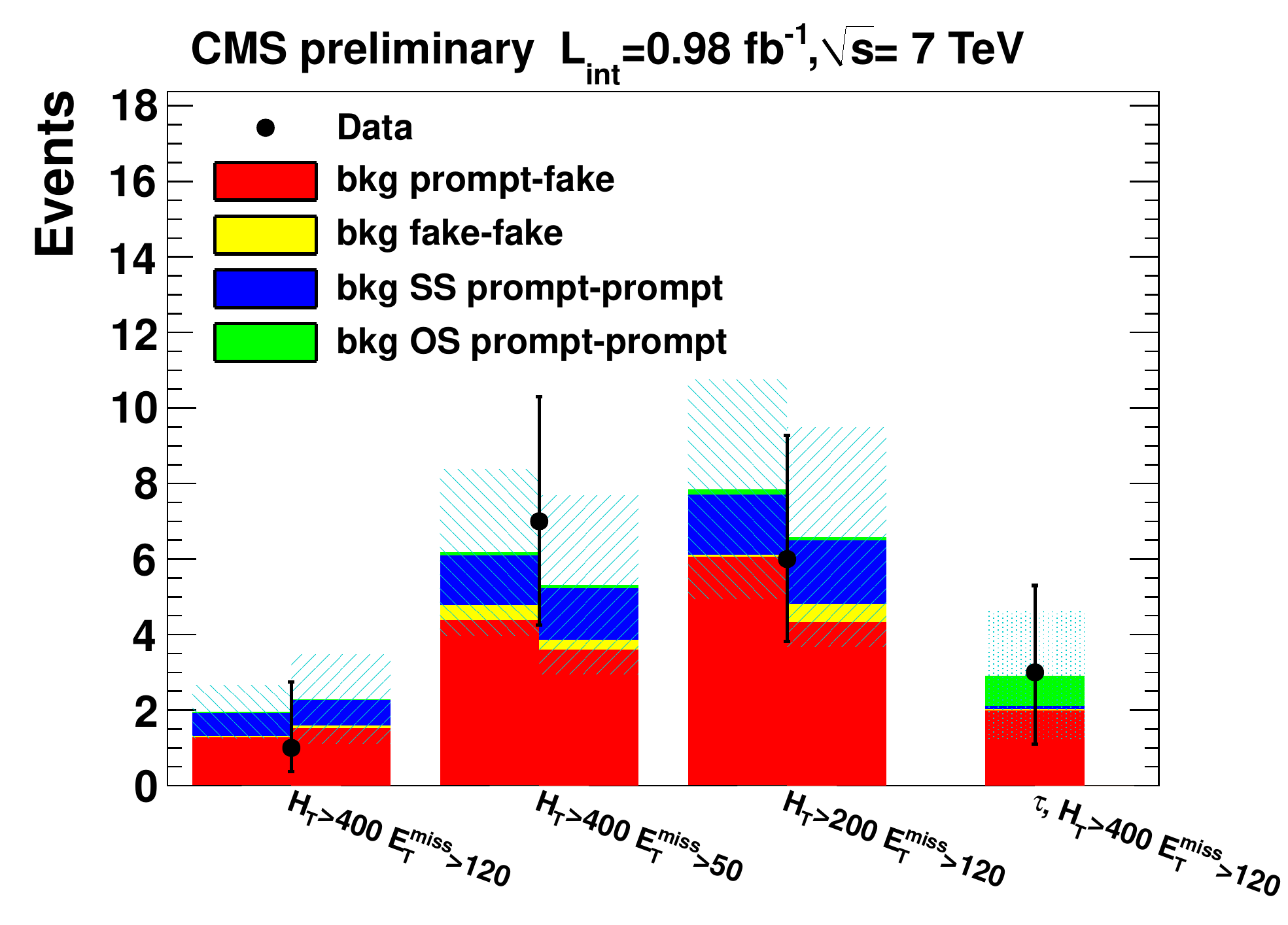}
\includegraphics[width=0.48\linewidth, height=0.33\linewidth]{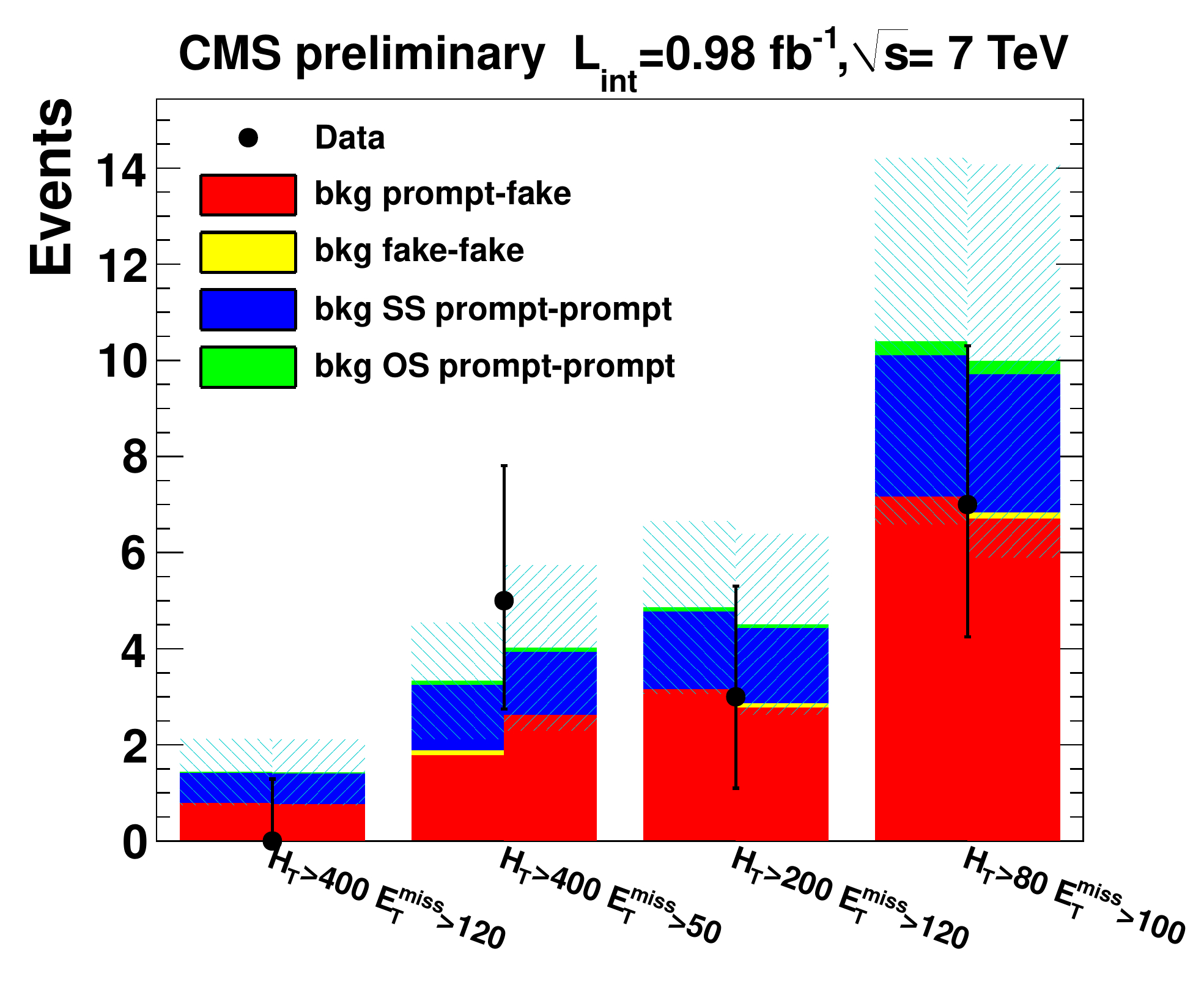}
\end{center}
\caption{Summary of background predictions and observed yields in the search regions for the $inclusive$ and $\tau$ (left), and $high$-$p_T$ $dileptons$ (right) selections.}
\label{fig:samesignplot}
\end{figure}

Search for new physics with same-sign isolated dileptons ($ee, e\mu, \mu\mu, e\tau, \mu\tau, \tau\tau$, where taus decay hadronically) with MET and hadronic 
jets are studied~\cite{bib:ss}. The baseline selection depending on the triggers used consists of $inclusive$ $dileptons$ ($H_T > 200$~GeV, MET > 30 GeV),  $high$-$p_T$ $dileptons$ 
($H_T > 80$~GeV, MET > 30 GeV) 
and $\tau$ $dileptons$ ($H_T > 350$~GeV, MET > 80 GeV). Electrons and muons with $p_T > 10$~GeV and $p_T > 5$~GeV respectively, are considered for $inclusive$ $dileptons$, 
whereas the $\tau$ selection uses $p_T > 15$~GeV. The $high$-$p_T$ $dileptons$ selection uses leptons with $p_T > 10$ GeV with the leading lepton above 20 GeV in the transverse momentum. 
The dominant SM background consists of events with jets mimicking a lepton signature (fake lepton) and charge mis-reconstruction 
(determined using $Z$ control samples). Contributions from events with fake lepton signatures that include heavy-flavor decays, electrons from unidentified photon conversions, 
hadrons reconstructed as leptons, etc. are estimated using ``tight-to-loose'' (TL) ratio. The $tight$ (leptons passing nominal selection) to $loose$ (leptons 
failing the nominal selection but passing a less restrictive selection) ratio is measured in data
in an independent sample dominated by QCD events. The total background is estimated by applying the TL ratio to a combination of double and single fake
lepton events in a given region. This method is used for the estimation of fake electrons, muons, and hadronic taus. Another complimentary method to 
estimate the fake lepton contribution uses
the measurement of the isolation distribution as a function of lepton $p_T$ and jet multiplicities in a sample enriched with $b\bar{b}$ production data. The 
isolation shape is then reweighted to the expected distribution in a simulated $t\bar{t}$ sample. SS dilepton backgrounds from decays of $W$, $Z$, referred to 
as ``prompt-prompt'' are estimated using MC simulations.
The results from these methods, charge flip contribution (for charge mis-reconstruction), as well as the prompt-prompt contribution for various signal search regions  
are shown in Fig.~\ref{fig:samesignplot}. The background expectations using two different data-driven methods agrees well with the observation. 
\section{Interpretation of results}
\begin{figure}[htb]
\begin{center}
\includegraphics[width=0.48\linewidth, height=0.33\linewidth]{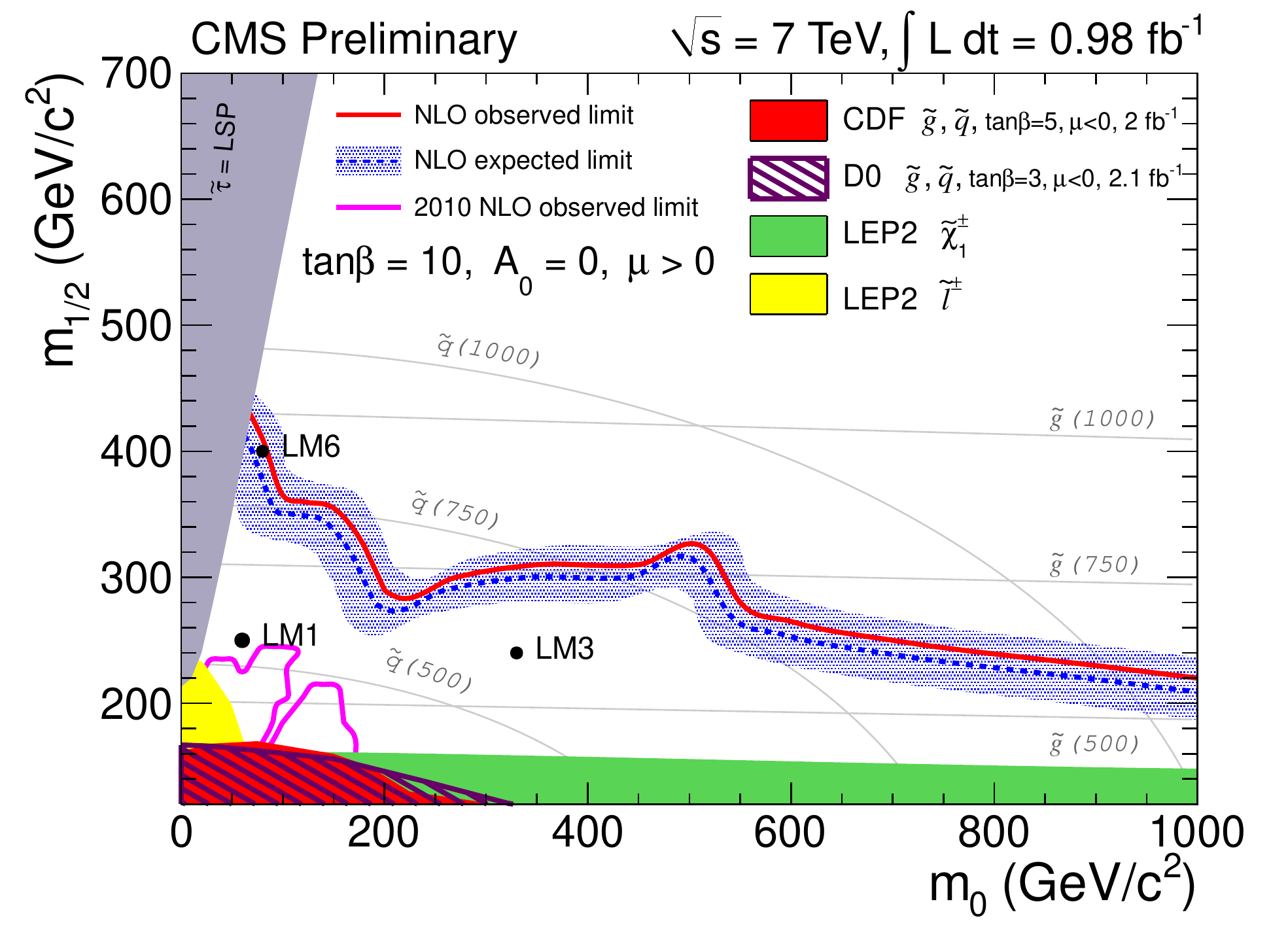}
\includegraphics[width=0.48\linewidth, height=0.33\linewidth]{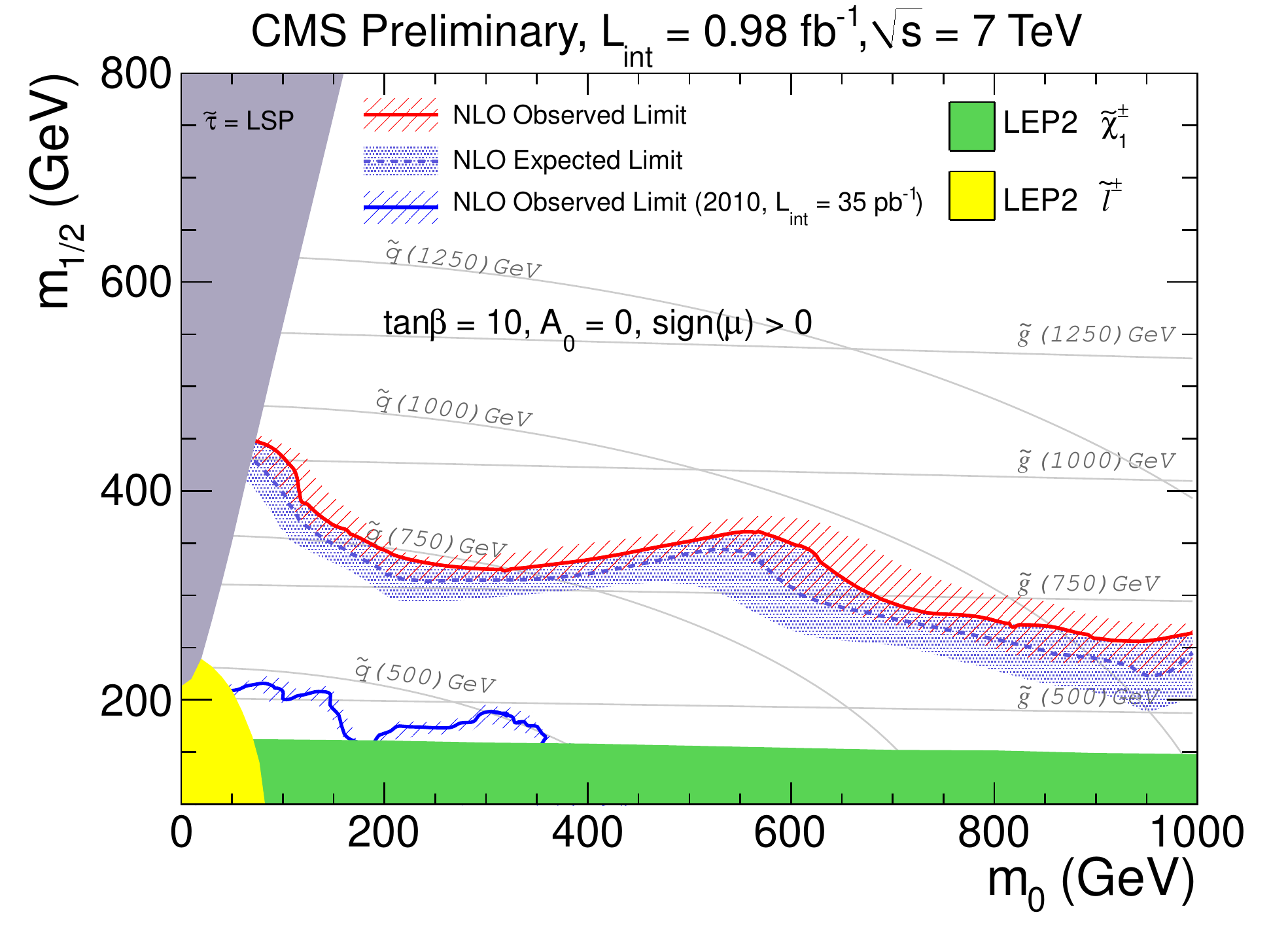}
\end{center}
\caption{Exclusion regions in the CMSSM using opposite-sign high $H_T$ (left) and same-sign $high$-$p_T$ $dileptons$ with $H_T > 400$~GeV, MET $> 120$~GeV (right) selections.}
\label{fig:msugraplot}
\end{figure}
Fig.~\ref{fig:msugraplot} shows the exclusion region in CMSSM framework for OS and SS searches as a function of $m_0$ and $m_{1/2}$ for tan$\beta = 10$, 
A$_0 = 0$ and $\mu > 0$. The dilepton search result extends to gluino masses of 825 GeV in the region with similar values of squark masses. 
\section{Summary}
Searches for Supersymmetry in final states with leptons are summarized using $\approx$ 1 fb$^{-1}$ of integrated luminosity. In all cases, dominant backgrounds 
are estimated from the data itself with minimal reliance on MC. In absence of any deviation from the SM, model independent upper limits are given 
along with interpretation in the CMSSM framework.

\end{document}